\documentclass[twocolumn]{aastex631}

\usepackage{xspace}

\newcommand\lya{Ly$\alpha$\xspace}{}
\newcommand\Lya{Ly$\alpha$\xspace}{}

\defcitealias{wyithe/dijkstra:2011}{WD11}

\usepackage{amsmath}

\newcommand{\angstrom}{\mbox{\normalfont\AA}}

\begin{document}

\title{Effect of Ly\ensuremath{\alpha} Radiative Transfer on Intensity Mapping Power Spectra}

\author[0000-0002-6907-8370]{Maja {Lujan Niemeyer}}
\affiliation{Max-Planck-Institut f\"{u}r Astrophysik, Karl-Schwarzschild-Str. 1, 85741 Garching, Germany}
\email{maja@mpa-garching.mpg.de}
 
\begin{abstract}
Clustering of Ly$\alpha$-emitting galaxies (LAEs) and Ly$\alpha$ line intensity mapping (LIM) are useful probes of cosmology. However, Ly$\alpha$ radiative transfer (RT) effects, such as absorption, line shift, and line broadening, and their dependence on the large-scale density and velocity fields can modify the measured LAE clustering and LIM statistics.
We explore the effects of RT on the Ly$\alpha$ LIM power spectrum in two ways: using an analytic description based on linear approximations, and using lognormal mocks. The qualitative effects of intergalactic Ly$\alpha$ absorption on the LIM auto- and cross-power spectrum include a scale-dependent, reduced effective bias, reduced mean intensity, and modified redshift-space distortions. The linear absorption model does not describe the results of the lognormal simulations well.
The random line shift suppresses the redshift-space power spectrum similar to the Fingers-of-God effect.
In cross-correlation of LAEs or Ly$\alpha$ intensity with a non-Ly$\alpha$ tracer, the Ly$\alpha$ line shift leads to a phase shift of the complex power spectrum, i.e. a cosine damping of the real part.
We study the impact of Ly$\alpha$ RT effects on the Hobby-Eberly Telescope Dark Energy Experiment (HETDEX) LAE and LIM power spectra using lognormal mocks. We find that even small amounts of IGM absorption will significantly change the measured LAE auto-power spectrum. The LAE-intensity cross-power spectrum stays within the measurement uncertainty. Therefore, HETDEX will be able to constrain Ly$\alpha$ RT effects.
\end{abstract}


\section{Introduction} \label{sec:intro}

The \lya emission line is an excellent tool for cosmology at high redshift \citep[e.g.,][]{partridge/peebles:1967}{}{}.
Detected \lya-emitting galaxies (LAEs) are used to measure their clustering 
and constrain cosmological parameters \citep[][]{ouchi/etal:2020,gebhardt/etal:2021}.
Instead of detecting individual LAEs in deep observations with high resolution, one can also map the total \Lya intensity in noisy, low-resolution observations to constrain cosmological parameters, called {line intensity mapping} \citep[LIM, e.g.,][]{bernal/kovetz:2022}{}{}.

Neutral hydrogen has a large scattering cross section around the \Lya line. The radiative transfer (RT) complicates the interpretation of measurements using \Lya emission. 
For example, using simulations, \citet{zheng/etal:2011} find a strong correlation between the \Lya optical depth in the intergalactic medium (IGM) and the large-scale density and velocity structure. They predict that the anisotropic dependence of the observed fraction of LAEs suppresses line-of-sight (LOS) density fluctuations, makes the effective bias scale-dependent, and can even `invert' the so-called {Kaiser effect} \citep[][]{kaiser:1987}, the linear redshift-space distortions (RSD). 

One can model the effect of IGM absorption on the LIM power spectra in various ways.
\citet{wyithe/dijkstra:2011} \citepalias[henceforth][]{wyithe/dijkstra:2011} and \citet{greig/komatsu:2013} derive an analytic model for this effect for the power spectrum and bispectrum. The analytic absorption model for the power spectrum in the first part of \citetalias{wyithe/dijkstra:2011} is based on linear approximations for the dependence of the \Lya transmittance on the matter, ionization rate, and velocity distributions. While this explains the qualitative effect of IGM absorption on the power spectrum, the amplitude is determined by three free parameters: the mean optical depth $\tau_0$, the fraction of \Lya photons subject to IGM absorption $F_\mathrm{abs}$, and the smoothing kernel of the ionization rate with respect to the galaxy distribution. 
The linear approximations are also only valid when the matter overdensity $\delta_\mathrm{m}$, the ionization rate perturbations $\delta_\Gamma$, and the velocity gradient perturbations $\delta_v$ are small, which is not the case in the immediate environments of galaxies. \citetalias{wyithe/dijkstra:2011} also present a more detailed analytic model that is based on assumptions on the density profile, ionization rate, temperature, and gas velocities in the environment of the LAEs.

Using a cosmological hydrodynamical simulation and post-processing it with \Lya RT, as done by \citet{zheng/etal:2011}, \citet{behrens/niemeyer:2013} and \citet{behrens/etal:2018}, may provide the most realistic estimate for the optical depth and the effect on the observed fluxes and the power spectrum, if it accurately simulates the matter and velocity structure within and outside of the galaxy halos.
However, the results of these simulations are dependent on the resolution: using an RT simulation with higher resolution, \citet{behrens/etal:2018} find little correlation between the large-scale environment and the observed fraction of LAEs, while they reproduce the results of \citet{zheng/etal:2011} when they degrade the resolution of the simulation. 

\citet{gurunglopez/etal:2020} develop a semi-analytic model for \Lya RT in the interstellar medium (ISM) and IGM. They find that, at low redshift ($z\in [2.2, 3]$), the spatial distribution of LAEs is independent of IGM properties. However, at $z=5.7$, the LOS velocity and density gradients modify the clustering of LAEs on large scales in an isotropic fashion.

Another effect of the \Lya RT in the ISM and circumgalactic medium (CGM) is the broadening and the shift of the \Lya line, typically toward the red \citep[e.g.,][]{nakajima/etal:2018}{}{}.
\citet{byrohl/saito/behrens:2019} find in a cosmological RT simulation that this wavelength shift is independent of the large-scale velocities, and they show that the shift adds a {Fingers-of-God}-like damping to the LAE power spectrum.

\citet[][]{croft/etal:2016}{}{} observationally find a strong effect of RT on the cross-correlation of quasars with \Lya intensity.
However, more recent studies are consistent with the absence of RT effects \citep[][]{croft/etal:2018,lin/etal:2022}{}{}.
More observations of LAE clustering or \Lya LIM are necessary to constrain RT effects.
The Hobby-Eberly Telescope Dark Energy Experiment \citep[HETDEX;][]{gebhardt/etal:2021}{}{} uses integral-field spectrographs to find $\simeq 10^6$ LAEs without target preselection in a $10.9\,\mathrm{Gpc}^3$ comoving volume. Its primary goal is to use LAE clustering statistics such as the power spectrum to constrain cosmological parameters, especially the dark energy equation of state \citep[][]{shoji/jeong/komatsu:2009}{}{}.
The blind nature of HETDEX enables \Lya LIM studies that may also be affected by \Lya RT effects. We therefore explore these effects on LIM power spectra and estimate the sensitivity of a HETDEX-like survey to these effects. Other \Lya LIM experiments \citep[e.g.,][]{dore/etal:2014,dore/etal:2016,dore/etal:2018,renard/etal:2021,renard/etal:2024} may also be affected by \Lya RT.

In this paper, we explore the effect of \Lya absorption in the IGM and the line shift and broadening on the \Lya intensity auto- and cross-power spectra. 
We use lognormal simulations for the forecast. 
\textsc{Simple}\footnote{\url{https://github.com/mlujnie/simple}} is a fast simulation tool for self-consistently generating galaxy catalogs and intensity maps in redshift space given an input power spectrum and luminosity function \citep[][]{lujanniemeyer/bernal/komatsu:2023}{}{}. It is based on a lognormal galaxy catalog generator\footnote{\url{https://bitbucket.org/komatsu5147/lognormal_galaxies}} \citep[][]{agrawal/etal:2017}{}{}, assigns luminosities, determines the detectability of each galaxy, and generates an intensity map. One can apply smoothing and a mask and add noise to make the mocks more realistic. Because the matter density and velocity fields are output by the lognormal galaxy simulations, one can self-consistently calculate the \Lya optical depth in each resolution element and attenuate the luminosities accordingly to simulate IGM absorption. Adding a random line shift and broadening is also straightforward.

A drawback of hydrodynamic simulations and semi-analytic models is their computational cost, which makes it unfeasible to generate enough realizations to calculate a covariance matrix and make sensitivity forecasts.
Because lognormal simulations are fast, it is possible to generate enough mocks to calculate the covariance matrix for the HETDEX survey and make predictions for its sensitivity to \Lya RT effects.

The rest of the paper is organized as follows. Section \ref{sec:igm_absorption_analytic} builds on the model by \citetalias{wyithe/dijkstra:2011} to develop an analytic model for the IGM absorption for LIM auto- and cross-power spectra. Section \ref{sec:Lya_line_shift_broadening} extends the work of \citet{byrohl/saito/behrens:2019} on the effect of the line shift and broadening of the \Lya line to LIM power spectra. Section \ref{sec:lognormal_simulation} describes the modifications to the \textsc{Simple} code to incorporate IGM absorption and a \Lya velocity shift, and shows the effects on the different power spectra using the lognormal simulations. Section \ref{sec:sensitivity_hetdex} analyzes the sensitivity of a HETDEX-like experiment to these effects using \textsc{Simple} mocks for the HETDEX survey. Section \ref{sec:discussion} discusses the shortcomings of this approach. We conclude in Section \ref{sec:summary_conclusion}.

 We use the following Fourier convention:
 \begin{equation}
 \begin{split}
     \tilde{f}(\mathbf{k}) &= \int \mathrm{d}^3\mathbf{x} f(\mathbf{x}) e^{i \mathbf{k} \cdot \mathbf{x}}\,, \\
     f(\mathbf{x}) &= \int \frac{\mathrm{d}^3 \mathbf{k}}{(2\pi)^3} \tilde{f}(\mathbf{k})e^{-i\mathbf{k} \cdot \mathbf{x}}\,,
\end{split}
\end{equation}
where the tilde denotes quantities in Fourier space. 
We refer to {real} space in contrast to {redshift} space,
and to {configuration} space in contrast to {Fourier} space.

Throughout this paper, we assume a flat $\Lambda$ cold dark matter ($\Lambda$\textsc{CDM}) cosmology with $H_0 = 67.66\,\mathrm{km\,s^{-1}\,Mpc^{-1}}$, $\Omega_{\mathrm{b},0}h^2 = 0.022$, $\Omega_{\mathrm{m},0}h^2=0.142$, $\ln\left(10^{10} A_s\right)=3.094$, and $n_s = 0.9645$.

\section{Intergalactic Ly\ensuremath{\alpha} Absorption}
\label{sec:igm_absorption_analytic}

A \Lya photon escaping the CGM of a galaxy that is close enough to the \Lya line center can scatter off of neutral hydrogen on the IGM. Although the photon scatters out of the LOS, we refer to it as absorption in this work.

The optical depth for a photon with the initial frequency $\nu$ on its way from a galaxy's virial radius to the observer is
\begin{equation}
\label{eq:tau_nu}
    \tau(\nu) = \int_{r_\mathrm{vir}}^\infty \mathrm{d}r\, n_\mathrm{H}(r) x_\mathrm{HI}(r) \sigma_\alpha\left[\nu\left(1+\frac{v_z(r)}{c}\right)\right],
\end{equation}
where $n_\mathrm{H}(r)$ is the hydrogen number density, $x_\mathrm{HI}(r)$ is the neutral fraction of hydrogen, $v_z(r)$ is the LOS velocity of the gas at distance $r$ from the galaxy, where $v_z>0$ if it moves away from the observer, and $\sigma_\alpha(\nu)$ is the \lya absorption cross section at frequency $\nu$ \citepalias[see, e.g.,][]{wyithe/dijkstra:2011}{}{}.
Using $x_\mathrm{HI} = n_\mathrm{H} \alpha_\mathrm{rec}^\mathrm{A} / \Gamma$ at photoionization equilibrium, 
where $\Gamma$ is the photoionization rate and $\alpha_\mathrm{rec}^\mathrm{A} \approx 4.18 \times 10^{-13}\,\mathrm{cm}^3\mathrm{s}^{-1}$ is the case-A recombination coefficient at temperature $T\approx 10^4\,\mathrm{K}$ \citep{burgess:1965,draine:2011}, we obtain
\begin{equation}
\label{eq:tau_1}
    \tau(\nu) = \int_{r_\mathrm{vir}}^\infty \mathrm{d}r\, \frac{n_\mathrm{H}^2(r)\alpha_\mathrm{rec}^\mathrm{A} \sigma_\alpha\left[\nu\left(1+\frac{v_z(r)}{c}\right)\right]}{\Gamma(r)}.
\end{equation}
Because the \lya cross section is within the integral, we can approximate it as a Dirac delta function:
\begin{equation}
\begin{split}
    \sigma_\alpha(\nu) &\approx f_\alpha \pi \frac{e^2}{m_e c} \delta(\nu-\nu_\alpha)\\
    &= f_\alpha \pi r_e c \delta(\nu-\nu_\alpha) =: \sigma_\alpha^\mathrm{tot}\delta(\nu-\nu_\alpha),
\end{split}
\end{equation}
where $\nu_\alpha$ is the \lya rest-frame frequency, $e$ is the electron charge, $m_e$ is the electron mass, $r_e = 2.81\times 10^{-13}\,\mathrm{cm}$ is the classical electron radius defined by $\frac{e^2}{r_e} = m_e c^2$, and $f_\alpha=0.4167$ is the oscillator strength of the \lya transition \citep[e.g., \citetalias{wyithe/dijkstra:2011}; ][]{Bartelmann_2021}{}{}.
Inserting this in equation \eqref{eq:tau_1} and integrating yields 
\begin{equation}
\label{eq:tau_deltafunc}
\tau_\delta = \frac{\left(c + v_z \right)
n_\mathrm{H}^2\alpha_\mathrm{rec}^\mathrm{A} \sigma_\alpha^\mathrm{tot}}
            { \nu_{\alpha} \Gamma \left| \frac{\mathrm{d}v_z^p}{\mathrm{d}r} {{-H(z)}}\right|},
\end{equation}
where $v_z^p$ denotes the peculiar velocity, such that $v_z = v_z^p - H(z)r$, and $H(z)$ is the Hubble parameter at redshift $z$.
Here, we used $\delta\left(g(r)\right) = \delta(r-r_0) \left| g^\prime(r_0) \right|^{-1}$, where $g^\prime$ denotes the derivative of $g$ with respect to $r$ and $g(r_0) = 0$. Specifically, $g(r) = \nu(r) - \nu_\alpha = \nu \left(1 + v_z(r)/c\right) - \nu_\alpha$.

Because equation \eqref{eq:tau_nu} only considers the absorption at distances larger than the virial radius, 
we summarize the RT within the galactic halo by using an effective absorption fraction. 
Like \citetalias{wyithe/dijkstra:2011}, we introduce the fraction $F_\mathrm{abs}$ of \lya photons subject to absorption as a free parameter,
such that the fraction $1-F_\mathrm{abs}$ of photons travels freely.
This is a simplified description of the line shape, where $1-F_\mathrm{abs}$ of the photons are redshifted outside of the high \Lya cross-section region from the RT within the ISM and CGM. The fraction $F_\mathrm{abs}$ of photons are either on the blue side of the \Lya line center or close enough to be subject to absorption. 
The total transmittance is given by integrating over the spectral flux density profile of the \lya line $J(\nu)$, which becomes
\begin{equation}
\label{eq:transmittance_from_tau}
    \mathcal{T} = \frac{\int \mathrm{d}\nu \, J(\nu) e^{-\tau(\nu)}}{\int \mathrm{d}\nu \, J(\nu)} \approx  1 - F_\mathrm{abs} + F_\mathrm{abs} e^{-\tau_\delta}.
\end{equation}

\subsection{Analytic Model for LIM Power Spectra}

We modify the calculation of \citetalias{wyithe/dijkstra:2011} to derive a model for intergalactic \lya absorption for the LIM power spectrum.
Consider the \lya intensity field as a biased tracer of the matter density with 
\begin{equation}
\delta I(\mathbf{x}) = I(\mathbf{x}) - I_0(\mathbf{x}) = I_0(\mathbf{x}) b_I \delta_\mathrm{m},
\end{equation} 
where $b_I$ is the linear intensity bias, $\delta_\mathrm{m} = \rho(\mathbf{x})/\rho_0(\mathbf{x}) - 1$
is the matter density contrast, and $\rho(\mathbf{x})$ is the matter density in real space.
The subscript $\empty_0$
denotes the mean field over many realizations -
for example, the mean intensity or matter density as a function of redshift.
We use brackets $\langle \cdot (\mathbf{x}) \rangle$ to denote the same when it is more convenient.
For simplicity, we consider a single redshift, and therefore $I_0(\mathbf{x}) = I_0 = \mathrm{const.}$
Thus, the intensity power spectrum is 
\begin{equation}
P_{II}(\mathbf{k}) = \langle | \widetilde{\delta I}(\mathbf{k}) |^2 \rangle 
=  b_I^2 I_0^2 P_\mathrm{m}(\mathbf{k}),
\end{equation}
where $P_\mathrm{m}$ is the matter power spectrum and we have neglected the discreteness of the intensity sources and therefore the shot-noise contribution.

The intensity after IGM absorption can be approximated as 
\begin{equation}
\label{eq:absorbed_intensity_fluctuations_1}
\begin{split}
    I^\mathrm{abs}(\mathbf{x}) &= I^{\mathrm{abs}}_{0} \left[1 + b_I \delta_\mathrm{m}(\mathbf{x})\right]\\
                   &+ \left[ \Gamma(\mathbf{x}) - \Gamma_0 \right] 
                   \left.\frac{\partial \mathcal{T}}{\partial \Gamma}\right|_{\mathcal{T}_0,\Gamma_0} 
                   \left.\frac{\partial I^{\mathrm{abs}}}{\partial \mathcal{T}}\right|_{\mathcal{T}_0}\\
                   &+ \left[ \rho(\mathbf{x}) - \rho_0 \right] 
                   \left.\frac{\partial \mathcal{T}}{\partial \rho}\right|_{\mathcal{T}_0,\rho_0} 
                   \left.\frac{\partial I^{\mathrm{abs}}}{\partial \mathcal{T}}\right|_{\mathcal{T}_0}\\
                   &+ \left[ \frac{\mathrm{d}v_z}{\mathrm{d}(a r_\mathrm{com})} - H \right] 
                   \left.\frac{\partial \mathcal{T}}{\partial \left(\frac{\mathrm{d}v_z}{\mathrm{d}(a r_\mathrm{com})}\right)}\right|_{\mathcal{T}_0,H} 
                   \left.\frac{\partial I^{\mathrm{abs}}}{\partial \mathcal{T}}\right|_{\mathcal{T}_0}.
\end{split}
\end{equation}
We adopt the linear model of \citetalias{wyithe/dijkstra:2011} for the transmittance:
\begin{equation}
\label{eq:transmittance_model}
\begin{split}
    \mathcal{T}(\delta_\mathrm{m},\delta_\Gamma,\delta_v) 
    &= (1-F_\mathrm{abs}) \\
    &+ F_\mathrm{abs} \exp\left\{ -\tau_0 \frac{1 + c_\gamma \delta_\mathrm{m}}{1 + \delta_\Gamma + \delta_v} \right\},
\end{split}
\end{equation}
where $\tau_0$ is the mean opacity in the IGM, $c_\gamma = (2.7-0.7\gamma) \simeq 1.72$ with the polytropic index $\gamma=1.4$  \citep{hui/gnedin:1997} denotes the dependence of the optical depth on dark matter density, and
$\delta_\Gamma = \frac{\Gamma}{\Gamma_0} - 1$ is the ionization rate perturbation.
The expression $\delta_v = \frac{\mathrm{d}v^p_z}{\mathrm{d}\left(a r_\mathrm{com}\right)}\frac{1}{H}$ represents the fluctuation in the LOS velocity.
The dependence of the intensity on the transmittance is 
\begin{equation}
\begin{split}
    \left.\frac{\partial I^{\mathrm{abs}}}{\partial \mathcal{T}}\right|_{\mathcal{T}_0} &= I_{0} = I_0^\mathrm{abs} \mathcal{T}_0^{-1} \\
    &\approx I_0^\mathrm{abs} \left( 1-F_\mathrm{abs} + F_\mathrm{abs} e^{-\tau_0}\right)^{-1}.
\end{split}
\end{equation}
We can rewrite equation \eqref{eq:absorbed_intensity_fluctuations_1}:
\begin{equation}
\begin{split}
    \delta I^\mathrm{abs}(\mathbf{x}) &= I^\mathrm{abs}(\mathbf{x}) - I^{\mathrm{abs}}_{0} \\
                                     &= I^\mathrm{abs}_0 \left( b_I \delta_\mathrm{m} + \delta_\mathrm{m} C_\rho 
                                        + \delta_\Gamma C_\Gamma + \delta_v C_v \right),
\end{split}
\end{equation}
where the constants $C_x$ are given by
\begin{equation}
\label{eq:c_values}
\begin{split}
    C_\Gamma &= \left.\frac{\partial \mathcal{T}}{\partial \mathrm{log}\,\Gamma}\right|_{\mathcal{T}_0,\Gamma_0} \left.\frac{\partial I^{\mathrm{abs}}}{\partial \mathcal{T}}\right|_{\mathcal{T}_0} (I_0^\mathrm{abs})^{-1}\\
    C_\rho &= \left.\frac{\partial \mathcal{T}}{\partial \mathrm{log}\,\rho}\right|_{\mathcal{T}_0,\rho_0} \left.\frac{\partial I^{\mathrm{abs}}}{\partial \mathcal{T}}\right|_{\mathcal{T}_0} (I_0^\mathrm{abs})^{-1}\\
    C_v &= \left.\frac{\partial \mathcal{T}}{\partial \mathrm{log}\left(\mathrm{d}v_z/\mathrm{d}r_\mathrm{com}\right)}\right|_{\mathcal{T}_0,H} \left.\frac{\partial I^{\mathrm{abs}}}{\partial \mathcal{T}}\right|_{\mathcal{T}_0} (I_0^\mathrm{abs})^{-1}.
\end{split}
\end{equation}
We find 
\begin{equation}
    C_\Gamma = C_v = \frac{F_\mathrm{abs} \tau_0 e^{-\tau_0}}{1 - F_\mathrm{abs} + F_\mathrm{abs} e^{-\tau_0}} =: C
\end{equation}
and 
\begin{equation}
    C_\rho = - \frac{c_\gamma F_\mathrm{abs} \tau_0 e^{-\tau_0}}{1 - F_\mathrm{abs} + F_\mathrm{abs} e^{-\tau_0}} = - c_\gamma C.
\end{equation}

The ionization rate fluctuations can be modeled by convolving the overdensity of ionizing sources with bias $b_\mathrm{ion}$
with a kernel $K_\lambda(k) = \mathrm{arctan}(k\lambda_\mathrm{mfp})/\left(k\lambda_\mathrm{mfp}\right)$, where $\lambda_\mathrm{mfp}$ is the mean free path of ionizing photons,
so that

\begin{equation}
\label{eq:ion_rate_smoothing_kernel}
    \tilde{\delta}_\Gamma (\mathbf{k}) = b_\mathrm{ion} \tilde{\delta}_\mathrm{m}(\mathbf{k}) K_\lambda(k).
\end{equation}

The fluctuations of intensity introduced by observing in redshift space, denoted by the superscript $^\mathrm{s}$, are
\begin{equation}
    \delta I^\mathrm{s} = \delta I - I_0 \frac{\mathrm{d}v_z}{\mathrm{d}r_\mathrm{com}}\frac{1}{Ha} = \delta I - I_0 \delta_v.
\end{equation}
Relating the velocity gradient fluctuations to the density fluctuations as $\tilde{\delta}_v(\mathbf{k}) = -f \mu^2 \tilde{\delta}_m(\mathbf{k})$,
where $f = \mathrm{d}\mathrm{ln} D / \mathrm{d}\mathrm{ln}a$ is the logarithmic growth factor,
we can write
\begin{equation}
\begin{split}
    \widetilde{\delta I^\mathrm{s}_\mathrm{abs}} (\mathbf{k}) &= I_0^\mathrm{abs} \tilde{\delta}_\mathrm{m}(\mathbf{k}) \\
    &\times \left[b_I + b_\mathrm{ion} K_\lambda(k) C_\Gamma + C_\rho + (1-C_v) f\mu^2\right] \\
    &=: I_0^\mathrm{abs} \tilde{\delta}_\mathrm{m}(\mathbf{k}) D_I(\mathbf{k}) = \mathcal{T}_\mathrm{g} I_0 \tilde{\delta}_\mathrm{m}(\mathbf{k}) D_I(\mathbf{k}).
\end{split}
\end{equation}
Here, we have assumed that the intrinsic luminosity of galaxies is uncorrelated with the local transmittance and $\mathcal{T}_\mathrm{g} = 1 - F_\mathrm{abs} + F_\mathrm{abs} {\int \mathrm{d}\mathbf{x}\, n(\mathbf{x}) e^{-\tau(\mathbf{x})}}/{\int \mathrm{d}\mathbf{x}\, n(\mathbf{x})}$ is the effective mean transmittance around galaxies.
The intensity auto-power spectrum is then given by  
\begin{equation}
\label{eq:analytic_power_spectrum_modified}
\begin{split}
    P_{II}(\mathbf{k}) = \langle |\widetilde{\delta I^\mathrm{s}_\mathrm{abs}}(\mathbf{k})|^2 \rangle = I_0^2 \mathcal{T}_\mathrm{g}^2 P_\mathrm{m}(k) D_I^2(\mathbf{k}).
\end{split}
\end{equation}

Taking a closer look at the intensity damping factor, we find
\begin{equation}
\begin{split}
    D_I &= b_I + b_\mathrm{ion} K_\lambda C - c_\gamma C + (1-C) f\mu^2\\
    &= \left( b_I + b_\mathrm{ion} C K_\lambda - c_\gamma C \right) \\
    &\times \left(1 + \frac{f}{b_I} \mu^2 \frac{b_I(1-C)}{b_I + b_\mathrm{ion}C K_\lambda - c_\gamma C}\right).
\end{split}  
\end{equation}
The RSD-like effect of the IGM absorption is introduced because the RSD parameter $f/b_I$ is effectively multiplied by the factor ${(1-C)}/({1 + C K_\lambda - c_\gamma C/b_I})$, assuming that $b_I = b_\mathrm{ion}$. This factor is smaller than one, i.e. the RSD is reduced, if $b_I > c_\gamma \simeq 1.72$ on small scales ($K_\lambda \simeq 0$), and $b_I > 0.5 c_\gamma \simeq 0.86$ on large scales ($K_\lambda \simeq 1$).

Following \citetalias{wyithe/dijkstra:2011}, the LAE overdensity in redshift space is
\begin{equation}
\label{eq:LAE_overdensity_FT_WD11}
\begin{split}
\widetilde{\delta}_\mathrm{g_\alpha}^\mathrm{s} &= \widetilde{\delta}_\mathrm{m} (\mathbf{k}) \left[ b_\mathrm{g_\alpha} + b_\mathrm{ion} C_\Gamma^\mathrm{g_\alpha} K_\lambda(k) + C_\rho^\mathrm{g_\alpha} + \left(1 - C_v^\mathrm{g_\alpha} \right) f \mu^2 \right]\\
&=: \widetilde{\delta}_\mathrm{m}(\mathbf{k}) D_\mathrm{g_\alpha}(\mathbf{k}),
\end{split}
\end{equation}
where 
\begin{equation}
\begin{split}
    C_\Gamma^\mathrm{g_\alpha} &= C_v^\mathrm{g_\alpha} = (\beta_\phi - 1) \frac{F_\mathrm{abs}\tau_0 e^{-\tau_0}}{1 - F_\mathrm{abs} + F_\mathrm{abs} e^{-\tau_0}} =: C^\mathrm{g_\alpha},\\
    C_\rho^\mathrm{g_\alpha} &= - c_\gamma C^\mathrm{g_\alpha},
\end{split}
\end{equation}
and $\beta_\phi>0$ is $-1$ times the slope of the \Lya luminosity function, which is also negative.
Note that, because $C_\rho^\mathrm{g_\alpha}$ is negative for $\beta_\phi > 1$, the effective LAE bias $b_\mathrm{g_\alpha} + b_\mathrm{ion} C_\Gamma^\mathrm{g_\alpha} K_\lambda(k) + C_\rho^\mathrm{g_\alpha}$ can become negative at large $k$ (where $K_\lambda(k)$ becomes negligible) if $c_\gamma C > b_\mathrm{g_\alpha}$;
for example, for $b_\mathrm{g_\alpha}\simeq 2$, $\beta_\phi \simeq 2.6$, and $\tau_0\simeq 1$.

The cross-power spectrum of LAEs and \Lya intensity is given by
\begin{equation}
\begin{split}
    P_{\mathrm{g_\alpha}\times I_\alpha} (\mathbf{k}) &= P_\mathrm{m}(k) I_0 \mathcal{T}_\mathrm{g} D_I(\mathbf{k}) D_\mathrm{g_\alpha}(\mathbf{k}).
\end{split}
\end{equation}
The cross-power spectrum becomes negative under the same conditions as $D_\mathrm{g_\alpha}(\mathbf{k})$.

For galaxies detected through a different line than \Lya that are not affected by \Lya RT effects, denoted by the subscript or superscript $g$, we have $\widetilde{\delta}_\mathrm{g} = \widetilde{\delta}_\mathrm{m}\left(b_\mathrm{g} + f\mu^2 \right)$, so that the cross-power spectrum of these galaxies with the \Lya intensity is 
\begin{equation}
    P_\mathrm{g\times I}(\mathbf{k}) = P_\mathrm{m}(k)\left(b_\mathrm{g} + f\mu^2\right)I_0 \mathcal{T}_\mathrm{g} D_I(\mathbf{k}).
\end{equation}

\begin{figure*}
\centering
\includegraphics[width=\textwidth]{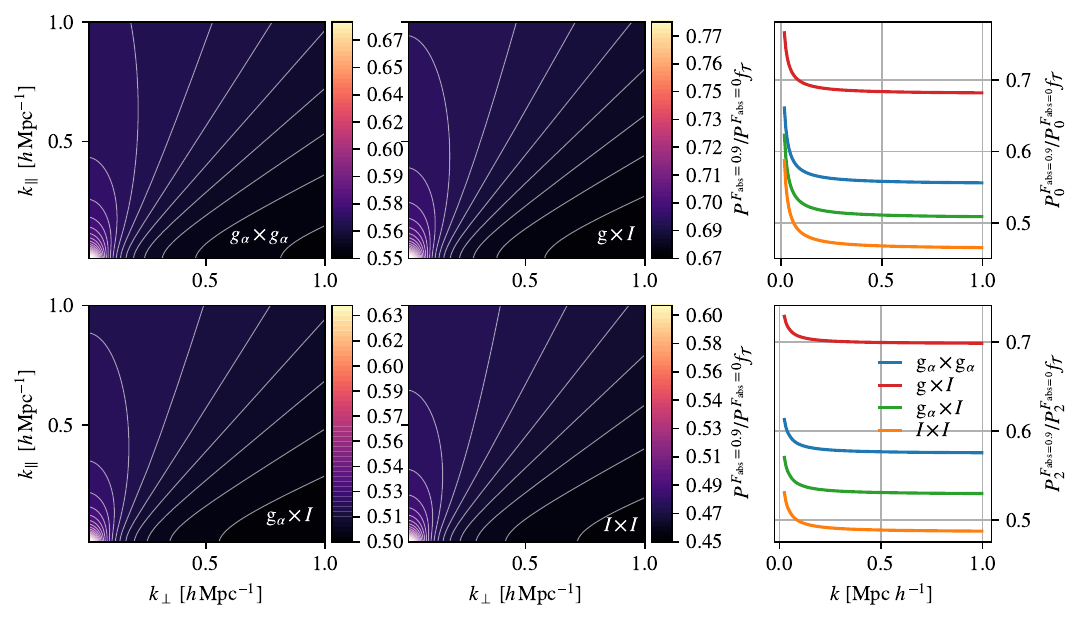}
\caption{Effect of IGM absorption on the \lya intensity power spectrum in the analytic model with $\tau_0 = 5$, $b_I = b_\mathrm{g_\alpha} = b_\mathrm{g} = 1.5$, $F_\mathrm{abs} = 0.9$, $\lambda_\mathrm{mfp} = 300\,\mathrm{Mpc}$, and $\beta_\phi = 1.8$.
We show $ P^{F_\mathrm{abs} = 0.9} / P^{F_\mathrm{abs} = 0.0} f_\mathcal{T}$, where $P^{F_\mathrm{abs} = 0.9}$ is the power spectrum with $F_\mathrm{abs} = 0.9$ and $P^{F_\mathrm{abs} = 0.0}$ is that without absorption. The factor $f_\mathcal{T}$ accounts for the amplitude change due to the lower mean intensity: $f_\mathcal{T} = I_0 / I_0^\mathrm{abs}$ for the cross-power spectra ($\mathrm{g}\times I$ and $\mathrm{g}_\alpha \times I$), $f_\mathcal{T} = (I_0 / I_0^\mathrm{abs})^2$ for the intensity auto-power spectrum ($I\times I$), and $f_\mathcal{T} = 1$ for the LAE auto-power spectrum ($\mathrm{g}_\alpha \times \mathrm{g}_\alpha$).
The four left and middle panels show the power spectrum damping as a function of wavenumber perpendicular and parallel to the LOS.
The top left panel shows the damping of the LAE auto-power spectrum, the top middle panel shows that of the cross-power spectrum of \Lya intensity with non-LAE galaxies. The bottom left panel shows the LAE-\Lya intensity cross-power spectrum damping. 
The bottom middle panel shows the \lya intensity auto-power spectrum. 
The top right panel shows the damping of the monopole of the LAE auto-power spectrum (blue), the non-LAE-intensity cross-power spectrum (red), the LAE-intensity cross-power spectrum (green), and the intensity auto-power spectrum (orange).
The bottom right panel shows the same for the quadrupoles.
Note that the anisotropy of the suppression depends strongly on the input parameters, especially the bias.}
\label{fig:Pk_par_perp_abs_effect}
\end{figure*}

\begin{figure*}
    \centering
    \includegraphics[width=\textwidth]{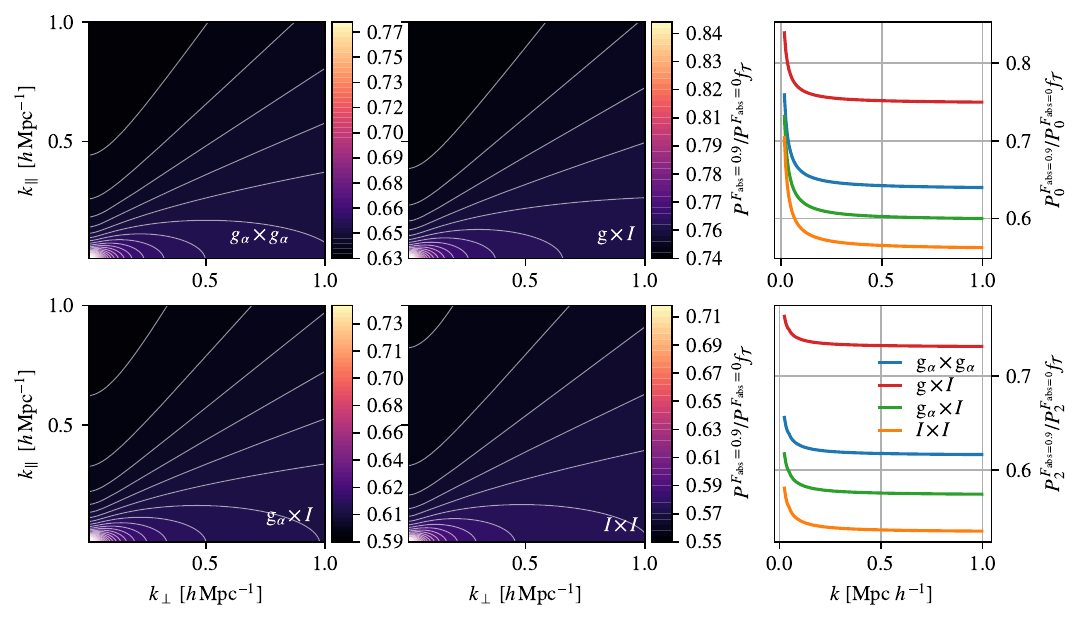}
    \caption{Same as figure \ref{fig:Pk_par_perp_abs_effect}, but with larger bias $b_I = b_\mathrm{g_\alpha} = b_\mathrm{g} = 2$.}
    \label{fig:Pk_par_perp_ratio_analytic_abs_effect_bias2}
\end{figure*}

Figures \ref{fig:Pk_par_perp_abs_effect} and \ref{fig:Pk_par_perp_ratio_analytic_abs_effect_bias2} show the effect of this model of \lya absorption in the IGM on the power spectra with different bias values. We set the mean optical depth to $\tau_0 = 5$ and the negative slope of the luminosity function to $\beta_\phi = 1.8$ in Figure \ref{fig:Pk_par_perp_abs_effect}, so that the parameters match those of the lognormal simulation in Section \ref{subsec:test_mock}. We set the mean free path of ionizing photons to $\lambda_\mathrm{mfp} = 300\,\mathrm{Mpc}$ \citep[][]{bolton/haehnelt:2007}{}{}.
The first-order effect of both settings is that the amplitude decreases when including IGM absorption because of the smaller effective bias. This amplitude difference does not include the lower mean intensity, which will further decrease the amplitude of the LIM power spectra. 
The reason for the smaller effective bias is that the transmittance modeled in equation \eqref{eq:transmittance_model} is smaller at higher densities. While a larger ionization rate implies a larger transmittance and the ionization rate is higher in high-density regions, its influence is reduced by the smoothing kernel.
The velocity gradient fluctuations are negative in overdensities, which also increases the transmittance.

The anisotropy of the suppression depends strongly on the input parameters.
In the configuration of Figure \ref{fig:Pk_par_perp_abs_effect} with the bias $b=1.5$, large scales are more strongly suppressed perpendicular to the LOS. A higher bias of $b=2$ inverts the RSD, leading to a stronger suppression along the LOS; see Figure \ref{fig:Pk_par_perp_ratio_analytic_abs_effect_bias2}.

The suppression of the monopole power spectrum shown in the right panels of Figures \ref{fig:Pk_par_perp_abs_effect} and \ref{fig:Pk_par_perp_ratio_analytic_abs_effect_bias2} shows that the effective bias is smaller at small scales than at large scales. This scale dependence is introduced by the smoothing kernel of the ionization rate parameterized by the mean free path of ionizing photons. A larger mean free path leads to a decrease at smaller $k \simeq \lambda_\mathrm{mfp}^{-1}$.

\subsection{Shot Noise}
We have ignored the shot-noise power spectrum in the previous section. Without absorption, and assuming constant redshift, the intensity auto-shot noise follows \citep[e.g.,][]{bernal/kovetz:2022}{}{}
\begin{equation}
\label{eq:shot_noise_noabs}
    P_\mathrm{shot} = \left(\frac{c}{4 \pi (1+z)^2 \lambda_\alpha H(z)} \right)^2 \int_{L_\mathrm{min}}^{L_\mathrm{max}} \mathrm{d}L\, \frac{\mathrm{d}n}{\mathrm{d}L} L^2,
\end{equation}
where $\lambda_\alpha$ is the rest-frame \Lya wavelength, $\frac{\mathrm{d}n}{\mathrm{d}L}$ is the luminosity function, and $L_\mathrm{min}$ and $L_\mathrm{max}$ are the minimum and maximum luminosities of the galaxies contributing to the intensity map.
The first factor assumes that we measure the specific intensity $I_\lambda$.
Assuming that the intrinsic luminosity of a galaxy is independent of the matter density and therefore uncorrelated with the local transmittance, equation \eqref{eq:shot_noise_noabs} turns into
\begin{equation}
    P_\mathrm{shot}^\mathrm{abs} = \left(\frac{c}{4 \pi (1+z)^2 \lambda_\alpha H(z)}\right)^2 \int_{L_\mathrm{min}^\prime}^{L_\mathrm{max}^\prime} \mathrm{d}L\, \frac{\mathrm{d}n}{\mathrm{d}L} L^2 \mathcal{T}_\mathrm{g}^2.
\end{equation}
If the galaxy sample changes - for example, because only undetected galaxies contribute to the intensity map through masking - the integration limits have to be changed: $L^\prime_\mathrm{min/max} = \mathcal{T}_g^{-1} L_\mathrm{min/max}$.

\section{Ly\ensuremath{\alpha} Line Shift and Broadening}
\label{sec:Lya_line_shift_broadening}

For \Lya photons to escape the ISM, they have to diffuse spatially and spectrally. In the absence of inflows or outflows, this gives rise to a symmetric, double-peaked spectrum, while simple shell models show that inflows enhance and outflows suppress the blue peak \citep[e.g.,][]{verhamme/schaerer/maselli:2006}{}{}. Because a sufficiently redshifted peak is redshifted out of the \Lya cross section, only the blue part of the spectrum is subject to intergalactic absorption. At redshifts $z \gtrsim 2$, LAEs predominantly have red peaks \citep[see, e.g.,][]{ouchi/etal:2020}{}{}.

The RT in the ISM also broadens the \lya line. We model the effect on the intensity auto- and cross-power spectra in the same way as spectral smoothing of the intensity map \citep[see, e.g., ][]{lujanniemeyer/bernal/komatsu:2023}{}{}, assuming that the broadening is independent of galaxy properties.
However, \citet{chung/etal:2021} show that the dependence of the line width on halo mass or luminosity produces a different damping of the power spectrum than a mass-independent broadening.
\citet{li/etal:2024} show that the LIM power spectrum is mostly sensitive to the line width, but not the exact line shape.
For modeling of the voxel intensity distribution including spectral broadening, see \citet{bernal:2024}.

When the redshift-space position of LAEs is determined from the \Lya line, it is affected by the line shift caused by RT as well as by the peculiar velocity of the galaxies. 
Following \citet{byrohl/saito/behrens:2019}, we consider the redshift-space galaxy density field that is exact under the assumption of one fixed global LOS direction $\hat{\mathbf{e}}_\parallel$ \citep[][]{taruya/nishimichi/saito:2010}{}{},
\begin{equation}
    \widetilde{\delta}_\mathrm{g}^s(\mathbf{k}) = \int \mathrm{d}^3\mathbf{x} \left[\delta_\mathrm{g}(\mathbf{x})- \partial_\parallel u_\parallel(\mathbf{x}) \right] e^{i\mathbf{k}\cdot \mathbf{x} + i k_\parallel u_\parallel(\mathbf{x})}.
\end{equation}
Here, we have introduced a scaled velocity $\mathbf{u} = \mathbf{v} / (a H)$ and $\partial_\parallel$ denotes the derivative with respect to the LOS distance. The same equation can be written for $\widetilde{\delta I^s}(\mathbf{k})$.
If we cross-correlate this galaxy overdensity with another field $\delta_\mathrm{g^\prime}$ that is not affected by $u_\parallel(\mathbf{x})$, and neglecting cross-shot noise, we can write the cross-power spectrum as
\begin{equation}
    P^s_{\mathrm{gg^\prime}}(\mathbf{k}) = \int \mathrm{d}^3\mathbf{r}\, e^{i\mathbf{k}\cdot \mathbf{r}} \langle e^{i k_\parallel u_\parallel(\mathbf{x})} \left[\delta_\mathrm{g}(\mathbf{x})- \partial_\parallel u_\parallel(\mathbf{x})\right] \delta_{\mathrm{g}^\prime}(\mathbf{x}^\prime) \rangle,
\end{equation}
where we have set the condition that the expression within the angle brackets depends only on $\mathbf{r} = \mathbf{x}^\prime - \mathbf{x}$.
We can rewrite this in terms of the cumulants as \citep[][]{scoccimarro:2004,taruya/nishimichi/saito:2010,byrohl/saito/behrens:2019}{}{}
\begin{equation}
\begin{split}
    P^s_{\mathrm{gg^\prime}}(\mathbf{k}) &= \int \mathrm{d}^3\mathbf{r}\, e^{i\mathbf{k}\cdot \mathbf{r}} \exp\left\{\langle e^{i k_\parallel u_\parallel(\mathbf{x})} \rangle_c \right\} \\
    &\times [ \langle e^{i k_\parallel u_\parallel(\mathbf{x})} A(\mathbf{x}) \delta_{\mathrm{g}^\prime}(\mathbf{x^\prime}) \rangle_c \\
    &+ \langle e^{i k_\parallel u_\parallel(\mathbf{x})} A(\mathbf{x}) \rangle_c \langle e^{i k_\parallel u_\parallel(\mathbf{x})} \delta_{\mathrm{g}^\prime}(\mathbf{x^\prime}) \rangle_c ],
\end{split}
\end{equation}
where $A(\mathbf{x}) = \delta_\mathrm{g}(\mathbf{x})- \partial_\parallel u_\parallel(\mathbf{x})$.
The factor $\exp\left\{\langle e^{i k_\parallel u_\parallel(\mathbf{x})} \rangle_c \right\} = \langle e^{i k_\parallel u_\parallel(\mathbf{x})} \rangle$ can be taken out of the integration because it does not depend on $\mathbf{r}$. It constitutes a {Fingers-of-God}-like damping of the form
\begin{equation}
    \begin{split}
D^\mathrm{cross}_{v_\mathrm{RT}} &= \exp\left\{\langle e^{i k_\parallel u_\parallel} \rangle_c \right\} = \langle e^{i k \mu u_\parallel} \rangle \\
 &= \int \mathrm{d}u_\parallel \phi(u_\parallel) e^{i k \mu u_\parallel},
\end{split}
\end{equation}
where $\phi(u_\parallel)$ is the probability density function (PDF) of the LOS velocity shift $u_\parallel$. This factor is a one-dimensional Fourier transform of $\phi(u_\parallel)$ to the variable $k \mu$.

As an example, consider a Gaussian PDF with mean $\bar{u}_\parallel$ and standard deviation $\sigma_{u_\parallel}$. The cross-power spectrum damping factor is then
\begin{equation}
\label{eq:cross_damping_vrt}
    D^\mathrm{cross}_{v_\mathrm{RT}} = e^{i k \mu \bar{u}_\parallel - \frac{1}{2}k^2\mu^2 \sigma_{u_\parallel}^2},
\end{equation}
which contains a phase shift due to $\bar{u}_\parallel$. The real component of the cross-power spectrum is therefore multiplied by 
\begin{equation}
\label{eq:D_fog_cross_real}
   \Re{\left(D^\mathrm{cross}_{v_\mathrm{RT}}\right)} = \mathrm{cos}(k \mu \bar{u}_\parallel) e^{- \frac{1}{2}k^2\mu^2 \sigma_{u_\parallel}^2}.
\end{equation}
The imaginary part of the power spectrum is multiplied by the respective sine function.
The cosine has a zero point at $k \mu = \pi / (2 \bar{u}_\parallel) \approx 0.9 \,h\mathrm{Mpc}^{-1}$ for $\bar{u}_\parallel = 200 \,\mathrm{km}\mathrm{s}^{-1}$ at $z = 3$.
Note that an auto-power spectrum of $\delta_\mathrm{g}^s$ will have a {Fingers-of-God}-like damping of the form
\begin{equation}
\label{eq:D_fog_auto}
   D_{v_\mathrm{RT}}^\mathrm{auto}(k, \mu) = \left| \int \mathrm{d}u_\parallel \phi(u_\parallel) e^{i k \mu u_\parallel} \right|^2,
\end{equation}
which is unaffected by $\bar{u}_\parallel$ \citep[see][]{byrohl/saito/behrens:2019}{}{}.

The phase shift can also occur in a cross-power spectrum of two fields with different velocity distributions, such as the cross-correlation between the detected, bright LAEs with the intensity of undetected, faint LAEs as planned by the HETDEX collaboration \citep[][]{lujanniemeyer/bernal/komatsu:2023}{}{}.

\section{Lognormal simulation}
\label{sec:lognormal_simulation}

\subsection{Modeling}
\label{subsec:modeling}

The analytic model is limited to the linear approximation of the optical depth in equation \eqref{eq:transmittance_model}, which is only expected to hold for small fluctuations in the matter density, ionization rate, and velocity. However, the \Lya absorption mostly happens in the immediate environment of the galaxies, where these fluctuations are large.

To introduce a model that is both fast and more accurate, we modify the \textsc{Simple} code \citep[][]{lujanniemeyer/bernal/komatsu:2023}{}{} to include the effect of \lya RT.
\textsc{Simple} is a tool for quickly generating mock intensity maps. 
It uses lognormal galaxy simulations \citep[][]{agrawal/etal:2017}{}{} and randomly assigns a luminosity to each galaxy by sampling from the input luminosity function. 
 One can smooth the map, add noise, and apply sky subtraction to make the mocks more realistic for observations. 
 One can also apply a selection function to obtain a catalog of detected galaxies.

The lognormal simulations of \citet{agrawal/etal:2017} calculate the velocity field
from the linearized continuity equation. 
Together with the matter density field and a model for ionization and the IGM transmittance, 
we can build a model for IGM absorption.

We model the ionization rate as proportional to the galaxy number density field of all (detected and undetected) galaxies, smoothed with the kernel $K_\lambda$ in equation \eqref{eq:ion_rate_smoothing_kernel}. The amplitude of the ionization rate is chosen so that the mean ionization rate matches that of \citet{khaire/srianand:2019} in each redshift bin. The mean free path of ionizing photons $\lambda_\mathrm{mfp}$ is left as a free parameter.

We calculate the hydrogen number density field $n_\mathrm{H}$ to be proportional to the matter density field:
\begin{equation}
n_\mathrm{H}(\mathbf{x}) = \frac{\rho_\mathrm{H}(\mathbf{x}) }{ m_\mathrm{P}} = \frac{0.75 \Omega_\mathrm{b}(z) {\rho}_{c}(z)}{m_\mathrm{P}} \left(1 + \delta_\mathrm{m}(\mathbf{x}) \right),
\end{equation}
where $\rho_\mathrm{H}$ is the hydrogen mass density, $m_\mathrm{P}$ is the proton mass, $\Omega_\mathrm{b}$ is the baryon density parameter, $\rho_c$ is the critical density, and $\delta_\mathrm{m}$ is the matter overdensity output from the lognormal simulation.

The velocity field is calculated by the lognormal simulation of \citet{agrawal/etal:2017}. 
We use \textit{numpy.gradient}\footnote{\url{https://numpy.org/doc/stable/reference/generated/numpy.gradient.html}} 
to calculate the velocity gradient.

Finally, we calculate the local optical depth $\tau_\delta(\mathbf{x})$ and transmittance $\mathcal{T}(\mathbf{x})$ in each cell with Equations \eqref{eq:tau_deltafunc} and \eqref{eq:transmittance_from_tau}. The luminosity $L(\mathbf{x})$ of each galaxy at the position $\mathbf{x}$ is replaced with $\mathcal{T}(\mathbf{x}) L(\mathbf{x}) = \left[1 - F_\mathrm{abs} + F_\mathrm{abs}e^{-\tau_\delta(\mathbf{x})} \right] L(\mathbf{x})$ before generating the intensity map. 
$F_\mathrm{abs}$ is equal for all galaxies.
We use this transmitted intensity map to calculate the power spectra and the transmitted flux for the selection function for the detected galaxy catalog.

Using a cosmological RT simulation, \citet{byrohl/saito/behrens:2019} find that the \lya velocity shift from RT is independent of the peculiar velocity of the host halo. We therefore model this effect by adding a random velocity shift to the mock galaxies following an input PDF $\phi(u_\parallel)$. 

Because line broadening can be modeled in the same way as a limited spectral resolution, one can increase the LOS smoothing in the input to \textsc{Simple}.

\subsection{RT Effects in Lognormal Simulations}
\label{subsec:test_mock}

We set up a cubic box with length $L_\mathrm{box} = 512\,\mathrm{Mpc}\,h^{-1}$ and $N_\mathrm{mesh} = 256$ at mean redshift $\bar{z}=2.2$ with galaxy bias $b = 1.5$ and the EWgt60 \Lya luminosity function of \citet{konno/etal:2016}, which includes photometrically selected LAEs with \Lya equivalent width larger than $60\,\angstrom$. We adopt a constant flux limit $F_\mathrm{min} = 3\times 10^{-18}\,\mathrm{erg}\,\mathrm{s}^{-1}\,\mathrm{cm}^{-2}$ for detection, no noise, and no smoothing of the intensity map. To remove the shot noise, we calculate the power spectrum using the {half-sum-half-difference} approach \citep[HSHD; see Appendix and][]{ando/etal:2018,wang/etal:2024}{}{}. We study the IGM absorption and the line shift effects separately. Realistically, line shift and broadening affect the amount of \Lya photons subject to absorption, which we include in $F_\mathrm{abs}$.

To exaggerate the IGM absorption effect, we adopt a large absorption fraction $F_\mathrm{abs} = 0.9$ and use all galaxies to generate the intensity map.
We set the mean free path of ionizing photons to $\lambda_\mathrm{mfp} = 300\,\mathrm{Mpc}$.

\begin{figure*}
    \centering
    \includegraphics[width=\textwidth]{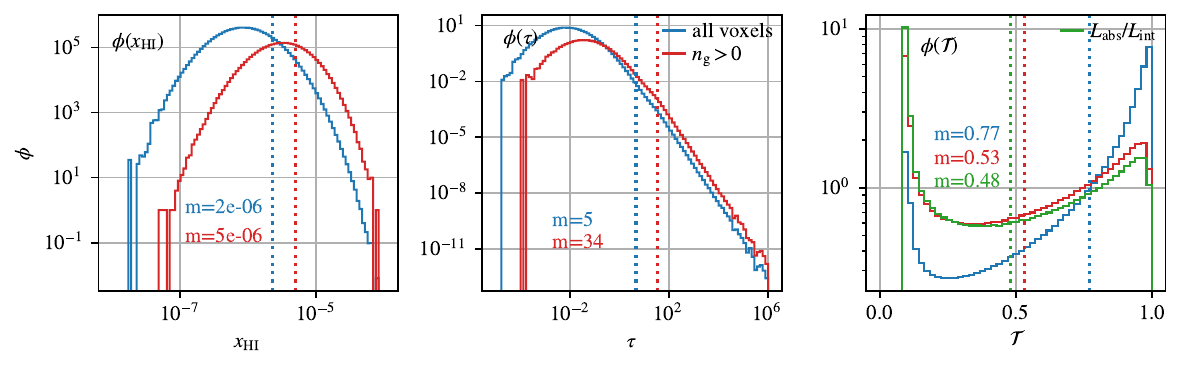}
    \caption{Distributions of the neutral hydrogen fraction $x_\mathrm{HI}$ (left panel), the optical depth (middle panel), and the effective transmittance with $F_\mathrm{abs} = 0.9$ (right panel) of the entire mock box (blue) of the test mock compared to those in voxels containing at least one galaxy (red).
    The green line in the right panel shows the effective transmittance of the galaxies calculated from ratio of the attenuated and original luminosity.
    The dotted lines and the text show the corresponding mean values.}
    \label{fig:transmittance_histogram_nonparamock}
\end{figure*}

Figure \ref{fig:transmittance_histogram_nonparamock} shows the distribution of the neutral hydrogen fraction, the optical depth, and the effective transmittance values (accounting for $F_\mathrm{abs}$) in the whole box compared to that in voxels containing galaxies and their mean values. The transmittance at galaxy positions is smaller than the overall mean transmittance in the simulation volume because galaxies lie in matter overdensities and therefore neutral hydrogen overdensities by construction.
The mean galaxy-weighted transmittance is low, $\bar{\mathcal{T}}_\mathrm{g}\simeq 0.5$.
The optical depth distribution has a long tail toward high optical depths. As a result, the mean optical depth is higher and inconsistent with the measurement of \citet{turner/etal:2024}, which is on the order of $\simeq 0.1$. The median optical depth in the lognormal simulations is lower at $\simeq 0.1$.

\begin{figure*}
    \centering
    \includegraphics[width=\textwidth]{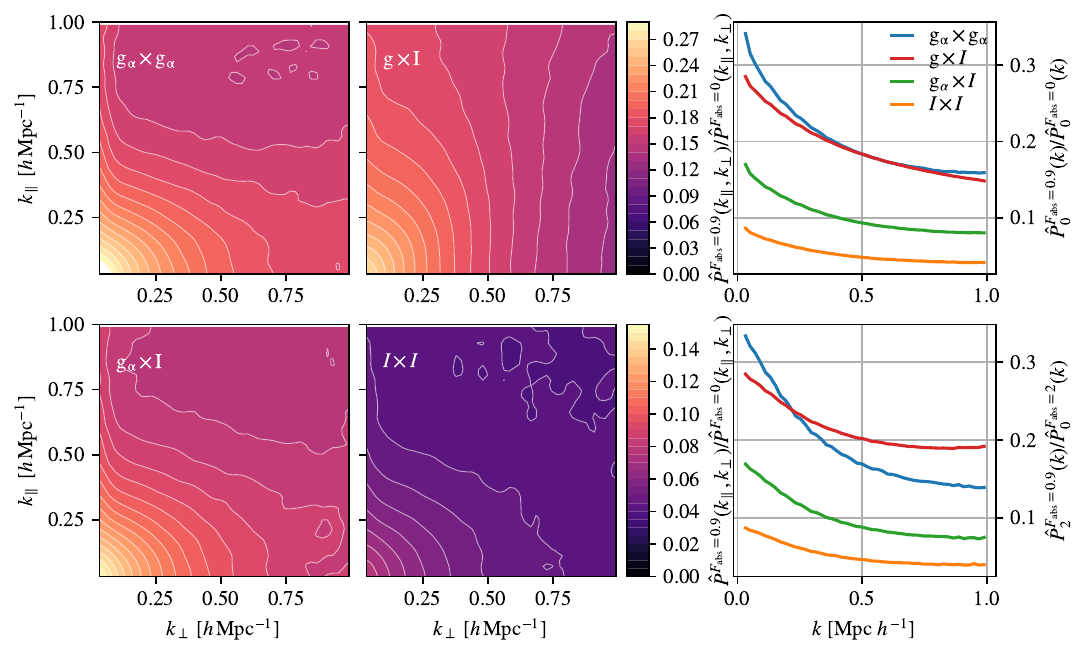}
    \caption{Similar to Figure \ref{fig:Pk_par_perp_abs_effect}, but for the lognormal mocks after shot-noise subtraction. These power spectrum ratios include the decreased mean intensity.
    The ratios of the power spectrum with absorption ($F_\mathrm{abs} = 0.9$) over that without absorption ($F_\mathrm{abs} = 0$) of the LAE auto-power spectrum (top left panel), the LAE-\Lya intensity cross-power spectrum (bottom left panel), the non-LAE-\Lya intensity cross-power spectrum (top middle panel), and the intensity auto-power spectrum (bottom middle panel) are shown. The right panels show the damping of the monopole and quadrupole of the LAE auto-power spectrum (blue), the non-LAE-\Lya intensity cross-power spectrum (red), the LAE-\Lya intensity cross-power spectrum (green), and the intensity auto-power spectrum (orange).
    The 2D damping maps were smoothed with a Gaussian kernel with a width of $\sigma \simeq 0.024\, h\mathrm{Mpc}^{-1}$ for better visualization.}
    \label{fig:Pk_par_perp_abs_effect_lognormal}
\end{figure*}

We calculate the intensity and LAE auto-power spectra, the LAE-\Lya intensity cross-power spectrum, and the cross-power spectrum of \Lya intensity with non-\Lya galaxies that have an uncorrelated luminosity function and are unaffected by IGM absorption. We subtract the shot noise using the HSHD method, and take the average power spectrum of $1000$ mocks.
Figure \ref{fig:Pk_par_perp_abs_effect_lognormal} shows the power spectrum ratios as a function of $k_\perp$ and $k_\parallel$ with and without IGM absorption. The main effect of the absorption is a suppression that is stronger at small scales. 
This is predicted by the analytic model, where the suppression is stronger at small scales where $K_\lambda$ is small.
However, the shape of the suppression differs from the analytic model in the setup with the same bias, luminosity function slope, and mean optical depth (see Figure \ref{fig:Pk_par_perp_abs_effect}). The suppression from IGM absorption in the lognormal simulations looks similar whether the bias is $b=2$ or $b=1.5$.

\begin{figure}
    \centering
    \includegraphics[width=0.47\textwidth]{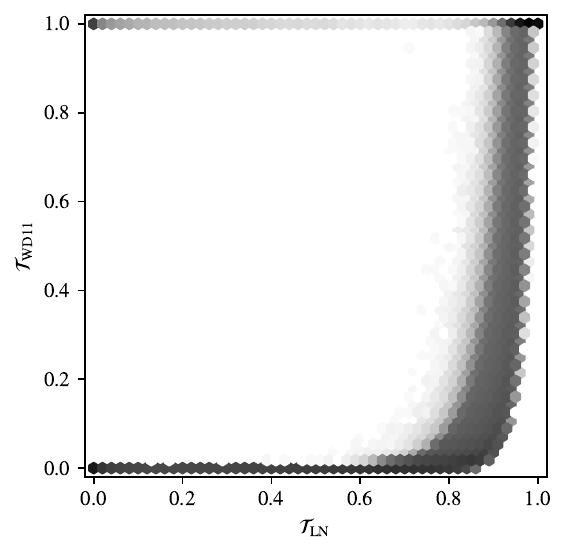}
    \caption{Transmittance calculated directly from the lognormal simulation $\mathcal{T}_\mathrm{
    LN}$ compared to that calculated with equation \eqref{eq:transmittance_model} using $\delta_\mathrm{m}$, $\delta_v$, $\delta_\Gamma$, and $\tau_0 = 5$ from the lognormal simulation ($\mathcal{T}_\mathrm{
    WD11}$), setting $F_\mathrm{abs}=1$ for both transmittance values. Darker regions contain more points, shown with a logarithmic color scale.}
    \label{fig:transmission_comparison}
\end{figure}

To explore the reason behind this discrepancy, we calculate the transmittance according to equation \eqref{eq:transmittance_model} using $\delta_\mathrm{m}$, $\delta_v$, $\delta_\Gamma$, and $\tau_0 = 5$ from the lognormal simulation. Figure \ref{fig:transmission_comparison} compares this transmittance to that directly calculated from the mocks.
It shows that the linear approximation for the optical depth in equation \eqref{eq:transmittance_model} does not describe the results of the lognormal simulations well. The absorption is dominated by the immediate surroundings of the galaxies, where the $\delta$ values are too large for linear approximations to hold.
However, lowering the value of $\tau_0$ in the WD11 model by a factor of 10 leads to a better agreement with the transmittance values of the lognormal simulation.

To model the \Lya line shift, we set $\phi(u_\parallel)$ to a Gaussian PDF with mean $\bar{v}_\mathrm{RT} = 639\,\mathrm{km}\,\mathrm{s}^{-1}$ and standard deviation $\sigma_{v_\mathrm{RT}} = 169\,\mathrm{km}\,\mathrm{s}^{-1}$. This line shift PDF is a best-fit Gaussian for the line shift distribution at $z=3.01$ with a galaxy number density $\bar{n}_\mathrm{g} = 10^{-3} \,\mathrm{Mpc}^{-3}h^3$ considering only the red peak in the RT simulation of \citet{byrohl/saito/behrens:2019}. In order to see the phase shift of the cross-power spectrum in this test, we keep the redshift-space positions of the galaxies in the galaxy catalog unchanged, while we add the line shift to the galaxies to calculate the intensity map. We use all galaxies to generate the intensity map in order to see the effect of the line shift on the cross-shot noise. 
We calculate the shot-noise-subtracted 2D power spectrum from an average of $100$ mocks using the HSHD method. We calculate the ratios between the power spectrum with and without the RT line shift and compute the mean damping along the LOS by averaging over $k_\perp$.
We confirm that the cross- and auto-power spectra follow the expected damping in Equations \eqref{eq:D_fog_cross_real} and \eqref{eq:D_fog_auto}.

\section{Sensitivity of a HETDEX-like Experiment}
\label{sec:sensitivity_hetdex}

We use the same HETDEX-like mocks as in \citet{lujanniemeyer/bernal/komatsu:2023} and include IGM absorption, a \Lya line shift, and \Lya line broadening to investigate the sensitivity of the power spectrum measured by a HETDEX-like survey to \Lya RT effects.
We set the mean free path of ionizing photons to $\lambda_\mathrm{mfp} = 300\,\mathrm{Mpc}$.
We set $\phi(u_\parallel)$ to a Gaussian PDF with mean $\bar{v}_\mathrm{RT} = 639\,\mathrm{km}\,\mathrm{s}^{-1}$ and standard deviation $\sigma_{v_\mathrm{RT}} = 169\,\mathrm{km}\,\mathrm{s}^{-1}$.
Fig. 13 of \citet{mentuchcooper/etal:2023} shows the observed line width distribution of the LAEs in HETDEX with a mean of $\sigma_{\lambda} = 3.54\,\angstrom$, which includes the intrinsic \lya line width of the LAEs and the smoothing of the spectrograph VIRUS \citep[$\sigma_\lambda \approx 2.38\,\angstrom$; see][]{hill/etal:2021}{}{}. To model the line broadening through RT and the VIRUS resolution, we apply Gaussian smoothing of the intensity map along the LOS with $\sigma_{\lambda} = 3.54\,\angstrom$ in the case with \Lya RT effects, and $\sigma_\lambda = 2.38\,\angstrom$ in the fiducial case without RT.
We subtract the shot noise using the HSHD method.

\begin{figure*}
    \centering
    \includegraphics[width=0.85\textwidth]{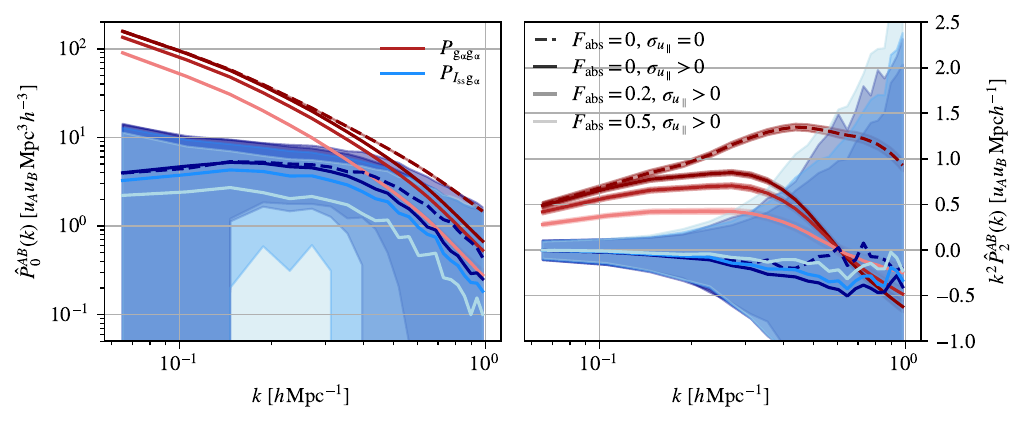}
    \caption{Power spectrum monopoles (left panel) and quadrupoles (right panel) with RT effects (solid lines) compared to the fiducial case without RT effects (dashed lines) in the $\bar{z} = 2.2$ bin. The red lines show the LAE auto-power spectra, and the blue lines show the cross-power spectra of LAEs with sky-subtracted intensity. The absorption fraction is shown with differently shaded lines, where lighter lines correspond to higher $F_\mathrm{abs}$. The shaded areas show the standard deviation of the mocks. The units are $u_\mathrm{g_\alpha} = 1$ and $u_I = 10^{-22}\,\mathrm{erg}\,\mathrm{s}^{-1}\,\mathrm{cm}^{-2}\,\mathrm{arcsec}^{-2}\,\angstrom^{-1}$.}
    \label{fig:HETDEX_absorption_pk0_pk2}
\end{figure*}

Figure \ref{fig:HETDEX_absorption_pk0_pk2} shows the impact of the RT effects on the HETDEX power spectra compared to the fiducial case in dashed lines at $\bar{z} = 2.2$\footnote{Note that, while the intensity unit contains $\mathrm{arcsec}^{-2}\,\angstrom^{-1}$, the intensity is not aperture-dependent. It is calculated using $I_\lambda(\mathbf{x}) = {c \rho_L(\mathbf{x})} / \left({4 \pi  (1+z)^2 \lambda_0  H(z)}\right)$, where $\rho_L(\mathbf{x})$ is the luminosity per unit comoving volume at location $\mathbf{x}$ and $\lambda_0$ is the rest-frame \Lya wavelength.}. We obtain a similar result for $\bar{z} = 3.0$. 
The fiducial galaxy auto-power spectrum quadrupole looks slightly different than in \citet{lujanniemeyer/bernal/komatsu:2023} because the HSHD method removes some previously unaccounted-for shot noise in the quadrupole. 
The LAE-\Lya intensity cross-power spectrum is smaller and has a lower signal-to-noise ratio because of a missing factor of $(1+z)^{-2}$ in the equation (4) for the specific intensity $I_\lambda$ in \citet{lujanniemeyer/bernal/komatsu:2023}. This is corrected in this paper and the public \textsc{Simple} code, such that the correct specific intensity is lower, while the noise level remains the same.

The amplitude, i.e., the effective bias, is lower for higher $F_\mathrm{abs}$. The RT line shift dispersion suppresses the power spectrum at small scales; this is also noticeable in the different shapes of the quadrupole. The effects are significant even at $F_\mathrm{abs} = 0.2$ for the LAE auto-power spectrum, while the LAE-\Lya intensity cross-power spectrum stays within the measurement uncertainty.
Note that the covariance of the power spectra with $F_\mathrm{abs}>0$ is overestimated because we do not change the input luminosity function, which is measured from observed fluxes, such that the number of observed galaxies is lower than for $F_\mathrm{abs} = 0$.

These results demonstrate that the HETDEX
LAE power spectrum is sensitive to the presence of \Lya RT effects. However, because the main effect of the \Lya absorption is degenerate with the LAE bias, it can be difficult to isolate - for example, through its scale dependence.
\citet{greig/komatsu:2013} show that the bispectrum can help break degeneracies between gravitational and RT effects. 
Using the power spectrum, HETDEX can nonetheless constrain the \Lya line shift distribution (see Section \ref{sec:Lya_line_shift_broadening}).
Because LAEs are mostly central halo galaxies and therefore unaffected by virial motion \citep[][]{ouchi/etal:2020}{}{}, a {Fingers-of-God}-like damping with a velocity dispersion of order $\simeq 100\,\mathrm{km\,s^{-1}}$ would likely stem from RT line shifts. 

\section{Discussion}
\label{sec:discussion}

Lognormal simulations directly produce the galaxy and matter distributions and linear velocities, which we use in this work to calculate the optical depth. This approach requires an assumption of the ionization rate smoothing kernel and the \Lya absorption fraction, but produces the mean optical depth as output. In this regard, there are fewer free parameters than in the analytic model. While the mean optical depth is dominated by a long tail toward high optical depths and inconsistent with the measurement of \citet{turner/etal:2024}, the median optical depth is lower and consistent with the measurement.

Because the lognormal simulations do not include galaxy-scale or CGM-scale physics, the optical depth is calculated from large-scale matter distributions and velocities, so a correlation between the optical depth and the galaxy distribution - and therefore an IGM absorption effect on the power spectrum - is inevitable.
We showed that this approach produces high optical depths in voxels containing galaxies. 
As in the \citetalias{wyithe/dijkstra:2011} model, the parameter $F_\mathrm{abs}$ defines how much of the \Lya RT takes place on the scale of the resolution of the simulations ($\simeq $Mpc), where $F_\mathrm{abs} = 0$ represents the case where no RT takes place on these scales.
This means that we account for the shape of the \Lya line emerging from the CGM only effectively with $F_\mathrm{abs}$ (see equation \eqref{eq:transmittance_from_tau}).

We approximate the RT outside of the virial radius as attenuation proportional to $\exp\left(-\tau_\delta\right)$. \citet{zheng/etal:2011} compare the clustering of LAEs in their full RT simulation to $\exp\left(-\tau(\nu)\right)$ attenuation from the centers of the galaxies. They find that, while the qualitative effects of RT on LAE clustering are captured by the $\exp\left(-\tau(\nu)\right)$ model, they are quantitatively different, presumably because multiple scatterings are unaccounted for in the $\exp\left(-\tau(\nu)\right)$ model.
One important difference between our model and that used by \citet{zheng/etal:2011} is that we calculate $\tau_\delta$ only outside of the virial radius, similarly to \citet{laursen/sommerlarsen/razoumov:2011}, and we approximate the RT within the galactic halo through the parameter $F_\mathrm{abs}$. As shown in Figure \ref{fig:transmittance_histogram_nonparamock}, most of the $\tau_\delta$ values in the lognormal simulations are smaller than $1$, where we expect multiple scatterings to be rare.

We find that the linear analytic model for absorption does not describe the lognormal simulations well. In the lognormal simulations, the absorption takes place in the immediate environment of the galaxies, where the matter overdensity is large. In this regime, the linear approximations for the transmittance and the effect on power spectrum break down. Therefore, the linear model is an inadequate description of the effect of \Lya absorption on the LIM power spectra.

A minor shortcoming of the lognormal approach is that we used the same luminosity function as an input, such that fewer galaxies are detected and the mean observed intensity is lower after IGM absorption. To mitigate this, one could change the input luminosity function to match it to the observed one after absorption. One could also implement a distribution of absorption fractions, which we assume would not change the power spectra, but it would slightly increase the covariance.

To summarize, the correlations between \Lya transmission and the large-scale matter and velocity, and therefore the effects of IGM absorption on the LAE and LIM power spectra, are model-dependent. Observations of the LAE and \Lya LIM power spectra, such as from HETDEX, are clearly necessary for constraints. As shown in this paper, HETDEX could be strongly affected by \Lya RT, and this must be accounted for in the power spectrum modeling.

In this work, we have only accounted for \Lya absorption in the IGM and \Lya line shifts and broadening from RT.
We have not attempted to study the impact of the \Lya absorption from background continuum sources on \Lya LIM power spectra. The \Lya forest in quasar spectra can easily be masked by masking quasar spectra for LIM.
However, \citet{weiss/etal:2024} find broad absorption troughs around LAEs in HETDEX through stacking, which will affect \Lya LIM studies.

This work also does not account for extended and diffuse \Lya emission, i.e. photons that scatter into the LOS or those that are produced in the CGM and IGM. Extended \Lya halos are ubiquitous around galaxies at $z\gtrsim 2$ \citep[e.g.][]{leclercq/etal:2017,lujanniemeyer/etal:2022a,lujanniemeyer/etal:2022b}. 
In this work, we assume that the flux originating from \Lya halos is included in the luminosity function and approximate the galaxies including their halos as point sources. Using high-resolution observations, \citet{leclercq/etal:2017} find that $\simeq 65\%$ of the LAE flux comes from \Lya halos. However, \citet{konno/etal:2016} use $2''-3''$ apertures to measure the \Lya flux of the LAEs for the luminosity function. The majority of the \Lya halo flux is contained within these distances \citep[][]{leclercq/etal:2017}.
\citet{lujanniemeyer/etal:2022a} find that \Lya halos of LAEs can reach $\simeq 160~\mathrm{kpc}$ (proper), but with negligible intensity compared to the central emission. The scales of interest in this work are also much larger than the extent of \Lya halos. Neither lognormal simulations nor the $\exp\left(-\tau_\delta\right)$ attenuation model are adequate for studying \Lya LIM at kpc scales.

Furthermore, \citet{byrohl/etal:2021} find that \Lya halo photons in the outer halo are scattered photons originating from galaxies outside of the host galaxy's dark matter halo, introducing a nonlocal component to \Lya halos. This environmental dependence of the intensity of the \Lya halo cannot be captured in this lognormal model.

\section{Summary and conclusion}
\label{sec:summary_conclusion}

We have presented an analytic model for the effect of \Lya absorption in the IGM on the \Lya LIM power spectra by adapting that of \citetalias{wyithe/dijkstra:2011}. While the overall effect is similar to that of the LAE auto-power spectrum - a lower, scale-dependent effective bias and reduced RSD, the suppression of the LIM component does not depend on the slope of the luminosity function. 

We have extended the model of \citet{byrohl/saito/behrens:2019} of the effect of line shifts from \Lya RT on the galaxy power spectrum to LIM power spectra. The effect on the intensity auto-power spectrum is the same as for LAEs. In cross-correlations of one tracer affected by this line shift with another that is unaffected, a phase shift of the power spectrum is introduced, leading to a cosine-shaped damping of the real part of the power spectrum. This can be useful to measure the average line shift and its dispersion of different galaxy populations with or without LIM.

We have modified the \textsc{Simple} code, a lognormal galaxy and intensity map simulator, to calculate the optical depth in each voxel from the matter and velocity distribution and attenuation of the intrinsic luminosities of galaxies in that voxel. 
In this model, a correlation between the optical depth and large-scale matter and velocity distributions is inevitable, but can be modulated with the effective absorption fraction $F_\mathrm{abs}$.
We also add a random RT line shift to the peculiar velocities. The analytic model for the line shifts matches the results of the lognormal simulations.

While both the analytic and the lognormal models for IGM absorption predict a stronger power spectrum suppression at small scales, their predictions for the dependence of the suppression on $k_\parallel$ and $k_\perp$ differ because the linear approximations break down in the nonlinear environment of the galaxies.

Finally, we have implemented the modified \textsc{Simple} code to model the effects of \Lya RT on the LAE and LAE-\Lya intensity power spectra for a HETDEX-like experiment. The line shift and broadening from RT significantly change the monopoles and quadrupoles of the LAE auto-power spectrum. The IGM absorption also changes the LAE auto-power spectrum significantly even at an absorption fraction of $0.2$, while the LAE-\Lya intensity cross-power spectrum remains within the measurement uncertainty even with an absorption fraction of $0.5$.
Therefore, HETDEX will help constrain the interplay of \Lya RT and galaxy clustering.
Our lognormal framework will be useful for the interpretation of upcoming large-scale structure measurements using \Lya emission.

\begin{acknowledgments}
I thank the anonymous referee for providing helpful comments.
I thank E. Komatsu for helpful discussions and comments on the draft. I acknowledge J. Niemeyer, C. Byrohl, and J. Bernal for insightful comments on the draft, and M. Gronke for interesting discussions. Thank you to T. Niemeyer and V. Niemeyer for their help with a calculation.

Computations were performed on the HPC system Freya at the Max Planck Computing and Data Facility.
\end{acknowledgments}

\vspace{5mm}

\software{astropy \citep{2013A&A...558A..33A,2018AJ....156..123A},  
          numpy \citep[][]{numpy_reference}{}{},
          simple \citep[][]{lujanniemeyer/bernal/komatsu:2023}{}{}}

\appendix
\section{Half-sum-half-difference Method for LIM}
\label{app:hshd_method_lim}

\citet{ando/etal:2018} introduce the {half-sum-half-difference} (HSHD) method for galaxy clustering to automatically remove shot noise. This is especially useful if the shot noise is anisotropic or scale-dependent.
One randomly splits the galaxy sample into two halves and calculates the density contrast for each, $\delta_{\mathrm{g}1}$ and $\delta_{\mathrm{g}2}$.
Then one calculates the half sum (HS) and half difference (HD) of the two fields:
\begin{equation}
    \mathrm{HS} = \frac{1}{2}\left(\delta_{\mathrm{g}1} + \delta_{\mathrm{g}2}\right); \; \mathrm{HD} = \frac{1}{2}\left(\delta_{\mathrm{g}1} - \delta_{\mathrm{g}2}\right).
\end{equation}
One can then calculate
\begin{equation}
\label{eq:HSHD_galaxies}
     \langle \widetilde{\mathrm{HS}}(\mathbf{k}) \, \widetilde{\mathrm{HS}}^\ast(\mathbf{k}^\prime) \rangle - \langle \widetilde{\mathrm{HD}}(\mathbf{k}) \, \widetilde{\mathrm{HD}}^\ast(\mathbf{k}^\prime) \rangle = \left(2\pi \right)^3 \delta_D(\mathbf{k} - \mathbf{k}^\prime) P^\mathrm{auto}_\mathrm{HSHD}(\mathbf{k}),
\end{equation}
where $\delta_D$ is the Dirac delta function.
The auto-power spectrum of HS contains the signal and shot noise, while that of HD only contains the shot noise, so that equation \eqref{eq:HSHD_galaxies} contains only the signal.
As shorthand, we use the notation 
\begin{equation}
    \hat{P}_\mathrm{HSHD}^\mathrm{auto} = \langle \mathrm{HS} \, \mathrm{HS}^\ast \rangle - \langle \mathrm{HD} \, \mathrm{HD}^\ast \rangle.
\end{equation}

\citet{wang/etal:2024} extend this method to the cross-power spectrum of two galaxy catalogs $A$ and $B$, where each galaxy in $A$ corresponds to a galaxy in $B$. One splits the catalogs into two halves so that $A_1$ and $B_1$ (and $A_2$ and $B_2$) maintain the one-to-one correspondence, and then calculates $\mathrm{HS}_{A/B}$ and $\mathrm{HD}_{A/B}$. The shot-noise-free cross-power spectrum estimator takes the form
\begin{equation}
    \hat{P}_\mathrm{HSHD}^\mathrm{cross} = \langle \mathrm{HS}_A \mathrm{HS}^\ast_B \rangle - \langle \mathrm{HD}_A \mathrm{HD}^\ast_B \rangle.
\end{equation}

LIM surveys typically have low resolution, such that the separation into two galaxy samples is not possible. After all, measuring the integrated emission from all galaxies within a resolution element is the main concept of LIM. 
Instead of separating galaxy samples, one can still separate observations of the same volume at different times and cross-correlate these to remove the intensity noise power spectrum.
The HETDEX survey, however, is not designed solely for LIM, but has a higher resolution in order to detect individual galaxies, which can be artificially decreased for LIM. In this case, it may be possible to separate galaxies within the same LIM voxel. 

For modeling purposes, it is helpful to separate the clustering power spectrum from the shot noise and intensity noise.
In the mock, we can randomly split the galaxies contributing to the intensity map into two separate samples and calculate $\delta I_{\mathrm{g}i} = I_{\mathrm{g}i} - \langle I_{\mathrm{g}i}\rangle$ for each sample $i$. 
Because only half of the galaxies contribute, the mean intensity is halved $\langle I_{\mathrm{g}i} \rangle = \frac{1}{2} \langle I_\mathrm{all} \rangle$.
Let us assume that $I_{\mathrm{g}i}$ includes uncorrelated intensity noise with the same variance $\langle I_\mathrm{noise}^2 \rangle = \sigma_\mathrm{I}^2$, so that $\langle \delta I_{\mathrm{g}i} \delta I^\ast_{\mathrm{g}j} \rangle = \langle \delta I_{\mathrm{g}i,\mathrm{signal}} \delta I^\ast_{\mathrm{g}j,\mathrm{signal}} \rangle + \sigma_\mathrm{I}^2 \delta_{ij}$, where $\delta_{ij}$ is the Kronecker delta.
Defining
\begin{equation}
    \mathrm{HS}_I = \delta I_{\mathrm{g}1} + \delta I_{\mathrm{g}2} ~ \mathrm{and} ~ \mathrm{HD}_I = \delta I_{\mathrm{g}1} - \delta I_{\mathrm{g}2},
\end{equation}
the power spectrum estimator becomes
\begin{equation}
\begin{split}
    \hat{P}_{II} = \langle \mathrm{HS}_I \mathrm{HS}^\ast_I \rangle - \langle \mathrm{HD}_I \mathrm{HD}^\ast_I \rangle
    = 2 \left(\langle \delta I_{\mathrm{g}1} \delta I^\ast_{\mathrm{g}2} \rangle + \langle \delta I_{\mathrm{g}2} \delta I^\ast_{\mathrm{g}1} \rangle \right) 
    = \frac{1}{2} \frac{\langle I_\mathrm{all} \rangle^2}{\langle I_{\mathrm{g}i} \rangle^2} \left( \langle \delta I_{\mathrm{g}1} \delta I^\ast_{\mathrm{g}2} \rangle + \langle \delta I_{\mathrm{g}2} \delta I^\ast_{\mathrm{g}1} \rangle \right).
\end{split}
\end{equation}
The power spectrum estimator has the same normalization $\langle I_\mathrm{all} \rangle^2$ as the ``normal'' power spectrum estimator $\langle \delta I_\mathrm{all} \delta I^\ast_\mathrm{all} \rangle$.
Note that $\mathrm{HS}_i$ and $\mathrm{HD}_i$ contain the intensity noise term $2 \langle \sigma_I^2 \rangle$, which cancels out in $\hat{P}_{II}$ as long as the intensity noise contributions of $\delta I_{\mathrm{g}1}$ and $\delta I_{\mathrm{g}2}$ are uncorrelated.

Similarly, the cross-power spectrum with a galaxy sample separated into $\delta_{\mathrm{g}1}$ and $\delta_{\mathrm{g}2}$ can be estimated as 
\begin{equation}
    \hat{P}_{I\mathrm{g}} = \langle \mathrm{HS}\,\mathrm{HS}^\ast_I \rangle - \langle \mathrm{HD}\,\mathrm{HD}^\ast_I \rangle,
\end{equation}
which does not contain shot noise as long as $\delta_{\mathrm{g}1}$ and $\delta I_{\mathrm{g}2}$ (and $\delta_{\mathrm{g}2}$ and $\delta I_{\mathrm{g}1}$) do not share galaxies contributing to both fields. This can be achieved by using the same galaxy split for $\delta_{\mathrm{g}i}$ and $\delta I_{\mathrm{g}i}$ in the mock.

The treatment of intensity noise and sky subtraction for the mocks is not straightforward. To ensure that the noise power spectrum is removed along with the shot noise, the noise maps of $\delta I_{\mathrm{g}1}$ and $\delta I_{\mathrm{g}2}$ must be uncorrelated. Depending on the split in the real data, the effective $\sigma_I$ of the subsamples is different than that of the total intensity. Staying agnostic, we add noise with the variance of the total intensity noise to each $\delta I_{\mathrm{g}i}$.
Because the galaxy split cannot be done after the sky subtraction in the mocks, we apply the sky subtraction to each $\delta I_{\mathrm{g}i}$. In a realistic setting, the sky is subtracted before splitting the data.

\bibliography{sample631}{}
\bibliographystyle{aasjournal}

\end{document}